\documentclass[aps,pra,twocolumn,groupedaddress,superscriptaddress,floatfix,amsmath,amssymb]{revtex4-1}

\usepackage{amsmath,amssymb}
\usepackage{graphicx}
\usepackage{epstopdf}
\usepackage{dcolumn}
\usepackage{bm}
\usepackage{braket}
\usepackage{times,txfonts}
\usepackage{hyperref}
\usepackage{color}

\begin{document}

\title{Estimating phase with a random generator: Strategies and resources in \\ multiparameter quantum metrology}

\author{Rozhin Yousefjani}
\email{r.yousefjani@sci.uok.ac.ir}
\affiliation{Department of Physics, University of Kurdistan, P.O.~Box 66177-15175, Sanandaj, Iran}

\author{Rosanna Nichols}
\email{rosanna.nichols@nottingham.ac.uk}
\affiliation{Centre for the Mathematical and Theoretical Physics of Quantum Non-Equilibrium Systems (CQNE), School of Mathematical Sciences, The University of Nottingham, University Park, Nottingham NG7 2RD, United Kingdom}

\author{Shahriar Salimi}
\email{shsalimi@uok.ac.ir}
\affiliation{Department of Physics, University of Kurdistan, P.O.~Box 66177-15175, Sanandaj, Iran}

\author{Gerardo Adesso}
\email{gerardo.adesso@nottingham.ac.uk}
\affiliation{Centre for the Mathematical and Theoretical Physics of Quantum Non-Equilibrium Systems (CQNE), School of Mathematical Sciences, The University of Nottingham, University Park, Nottingham NG7 2RD, United Kingdom}

\date{\today}

\begin{abstract}

Quantum metrology aims to exploit quantum phenomena to overcome classical limitations in the estimation of  relevant parameters. We consider a probe undergoing a phase shift $\varphi$ whose generator is randomly sampled according to a distribution with unknown concentration  $\kappa$, which introduces a physical source of noise. We then investigate strategies for the joint estimation of the two parameters $\varphi$ and $\kappa$ given a finite number $N$ of interactions with the phase imprinting channel. We consider both single qubit and multipartite entangled probes, and identify regions of the parameters where simultaneous estimation is advantageous, resulting in up to a twofold reduction in resources. Quantum enhanced precision is achievable at moderate $N$, while for sufficiently large $N$ classical strategies take over and the precision follows the standard quantum limit. We show that full-scale entanglement is not needed to reach such an enhancement, as efficient strategies using significantly fewer qubits in a scheme interpolating between the conventional sequential and parallel metrological schemes yield the same effective performance. These results may have relevant applications in optimization of sensing technologies.
\end{abstract}

\maketitle

\section{Introduction}\label{sec:Introduction}

Quantum metrology aims to realize measurements with a precision beyond any threshold achievable by using classical methods alone \cite{Giovannetti2006}. To reach this goal, quantum phenomena need to be suitably exploited as resources \cite{Braun,Pezze}. This field of research is of pivotal importance for the improvement of fundamental metrological standards, enabling widespread applications such as in timing, healthcare, defence, navigation, and astronomy \cite{Giovannetti,navigation,biology,gravwaves}.

Measurements of physical quantities, such as the strength of a field, a force, displacements, changes in concentration or time, can very often be recast in terms of a {\it phase estimation} scheme \cite{Giovannetti,like phase estimation}. This scheme is paradigmatic in displaying the quantum enhancement possible in metrology. Such an advantage is  normally defined by the reduced scaling of the error on the estimated parameter as a function of the number of interactions, $N$, between the adopted probe and the channel imprinting an unknown phase $\varphi$ on it. It is customary to refer to an inverse linear scaling of the variance $\delta^2 \varphi \sim N^{-1}$ as to the standard quantum limit (SQL), which is just a consequence of classical statistics in the regime of a large number $N$ of repetitions. Conversely, exploiting quantum resources such as coherence (manifested as asymmetry with respect to the phase generator) \cite{Braun,coherence} or multipartite entanglement \cite{Pezze}, it is ideally possible to reduce the error on the estimate of $\varphi$ by a quadratic factor, reaching the so-called Heisenberg limit $\delta^2\varphi \sim N^{-2}$.

However, a central aspect of phase sensing in a real-world scenario is the interaction between the probing system and the environment. Unfortunately, when this is taken into account, the promised quantum advantage is severely affected, with the enhancement becoming, at best, a constant factor in the asymptotic limit of large $N$, under most commonly encountered noise models  \cite{noisy freqency estimation,Yousefjani,lossy interferommetry,macconehierarchy}. This motivates one to examine more carefully the practical regime of finite $N$, where some form of advantage may remain \cite{Rosanna,Yousefjani,Braun}.

In many of the past studies on noisy quantum metrology, the analyzed task was purely that of phase (or frequency) estimation, under the assumption of knowing the details of the noise sources e.g.~by means of prior information. However, one may also want to directly estimate the noise on the system, which may in turn allow for an improved estimation of the phase parameter, as well as being of interest in its own right. Rather than estimating the noise and phase parameters individually, some advantage may be attained by simultaneous estimation, bringing us into the field of {\it multiparameter quantum metrology} \cite{Fisher inf. Matrix,Review,Ragy}.

From an experimental point of view, estimating parameters simultaneously in a single metrology protocol can be more challenging than estimating each of them individually. However, in the best case scenario, considering an estimation of, say, $p$ parameters at once, the measurement precision of one parameter will be totally unaffected by the simultaneous measurement of the others, reducing the resources required by a factor of $p$. This is known as the parameters being \emph{compatible} \cite{Ragy}. In general, exploring problems and developing solutions for multiparameter quantum metrology may not only result in an advantage in high-level applications such as
microscopy, spectroscopy, optical or magnetic field sensing, or gravitational wave detection \cite{magfields,gravwaves}, but also provide deeper insights on multipartite quantum correlated states and quantum measurements. Partly motivated by these perspectives, a number of recent works have investigated compatibility and simultaneous estimation of two or more parameters \cite{Review,Liu,TsangX,Ragy}, be they associated to noise characteristics \cite{two noise}, or multiple phases \cite{unitary parameters}, or some instances of phase and noise  \cite{unitary and noise}.

In this work we look at multiparameter estimation in the context of a noise model where the phase imprinting operation on qubit probes is realized by a unitary with a randomly sampled generator. This provides insight into multiparameter quantum metrology more broadly, as an example of estimation under a non-trivial noise model, and a scheme where an enhancement is attainable even when it is lost in the asymptotic limit and when the parameters are not always fully compatible. The model proves particularly instructive as the noise can be completely defined by a single parameter $\kappa$, corresponding to the inverse width of the distribution of the phase generator \cite{Rosanna}, so that our analysis can more easily outline definitive metrological strategies for the joint estimation of phase ($\varphi$) and noise ($\kappa$), and single out an efficient use of quantum resources to fulfill these strategies.

\begin{figure}[bt]
	\includegraphics[width=8.5cm]{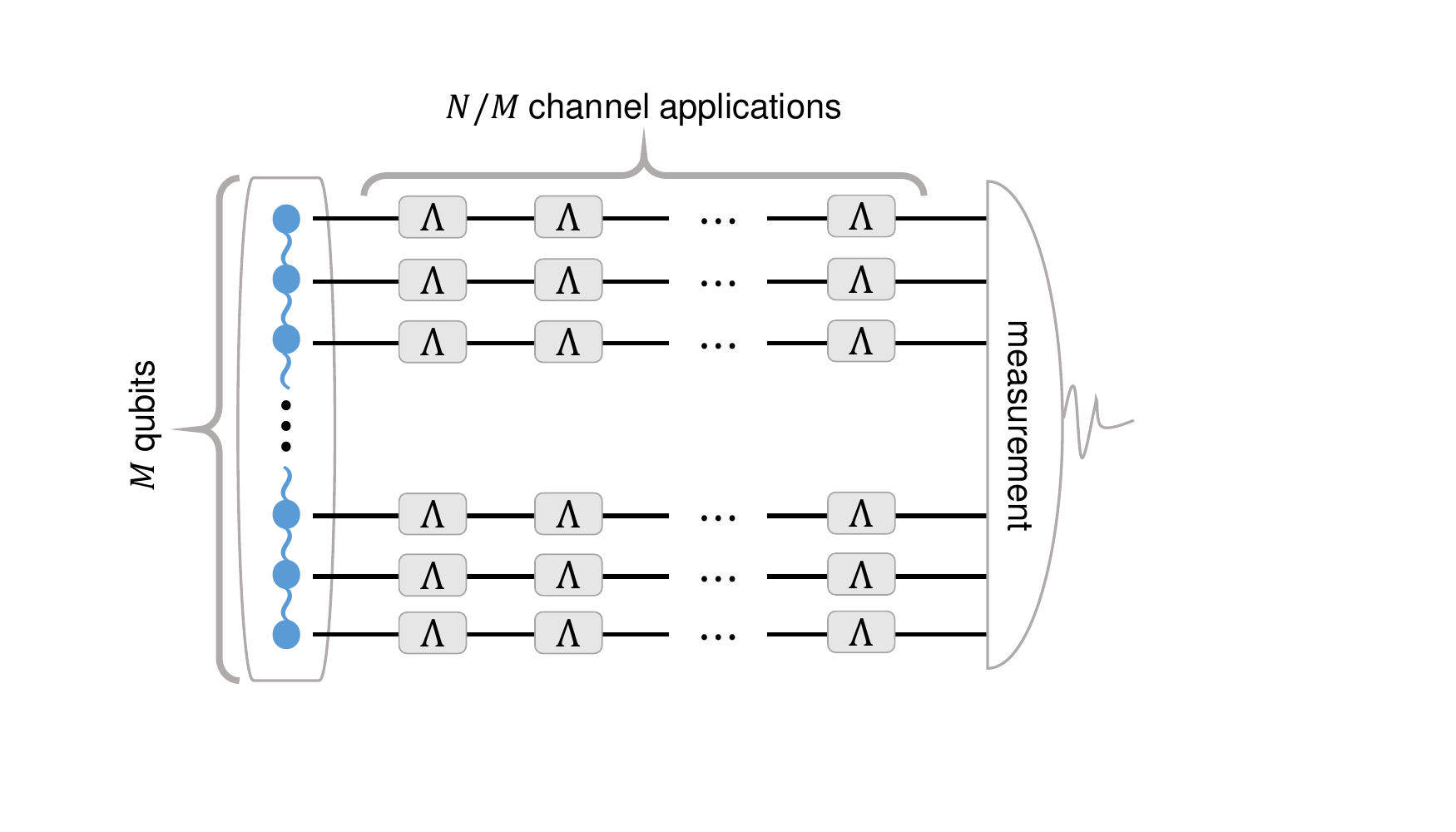}
	\caption{A general scheme for quantum metrology. Each qubit of an $M$-partite  entangled state is subjected to a sequence of $N/M$ applications of a channel $\Lambda$, which imprints the parameters to be estimated onto the probe. For $M=1$ this reduces to a sequential scheme relying on single qubit coherence and $N$ iterations of the channel, while for $M=N$ this reduces to a fully parallel scheme relying on $N$-qubit entanglement and a single application of the channel on each qubit. All the plotted quantities are dimensionless.}
	\label{interpolate}
\end{figure}

The paper is organized as follows. In Section~\ref{sec:QuantumParameterEst} we recall the basics of multiparameter quantum estimation theory,
 while the noise model under consideration is described in Section~\ref{sec:Noise}. In Sections~\ref{sec:1qubit} and \ref{sec:2qubit} we first investigate in detail the estimation of $\varphi$ and $\kappa$ using single qubit and two qubit probes, respectively, studying the optimal initial probe states and the compatibility. We find that, although the parameters are not always fully compatible, the simultaneous estimation scheme is always superior, with the degree of this superiority being dependent on the values of the parameters to be estimated. In Section~\ref{sec:Nqubit} we then move to the case of using $N$-qubit parallel entangled probes at the input. Here, we define the metrological strategies that arise and the areas of the parameter space where each strategy is relevant. We find that there is always a quantum enhancement at finite $N$ before the quantum schemes are overtaken by their classical counterparts at high $N$. However, whether individual or simultaneous estimation is optimal relies both on the parameter values and the size of the probe available. Finally, in Section~\ref{sec:roleofM} we explore a more general scheme for quantum metrology (see Figure~\ref{interpolate}), a hybrid of the parallel and sequential metrological schemes usually considered \cite{Giovannetti2006}, in which each qubit of an $M$-qubit entangled probe is passed through the phase imprinting channel $N/M$ times such that the total number of applications, $N$, remains constant. It is known that parallel, $M=N$, schemes are typically more robust under noise than sequential, $M=1$, schemes \cite{macconehierarchy}, but are far more difficult to implement experimentally. It is therefore striking that we find the error saturates very quickly with $M$, that is, an entangled probe state of only a few particles may achieve an error arbitrarily close to that of a large-scale multipartite entangled probe. This observation is independent of the parameter values or the total number of channel applications and may aid the transition to experimentally feasible quantum sensing and metrology.

\section{Quantum parameter estimation} \label{sec:QuantumParameterEst}
A typical estimation protocol consists of the following steps:
First, a quantum probe state, $\rho_{0}$, is prepared. Next, the probe state is modified by some physical mechanism, encoding the set of parameters to be estimated, $\{\mu\}$, onto the state.
Formally, this encoding can be described by a parameterized completely positive trace preserving (CPTP) map, $\Lambda_{\{\mu\}}$.
Finally, the evolved probe state, $\rho_{\{\mu\}}=\Lambda_{\{\mu\}}[\rho_{0}]$, is measured and estimates of the parameters are obtained through classical post-processing of the measurement results.
The performance of an estimator can be quantified by the covariance matrix, $\mathrm{Cov}\left(\rho_{\{\mu\}}\right)$, which captures both the variance of---and therefore the error on---each of the individual parameters, as well as the covariance---and therefore some indication of the correlation---between them.
For unbiased estimators, the quantum Cram\'{e}r-Rao bound establishes a lower bound to the covariance matrix in terms of the quantum Fisher information matrix (QFIM) \cite{Fisher inf. Matrix}
\begin{equation}\label{QCRB}
\mathrm{Cov}\left(\rho_{\{\mu\}}\right)\geq \mathcal{F}^{-1}.
\end{equation}
The QFIM $\cal F$ contains information on the parameters of a system that is acquired by performing an optimal measurement on it.
The entries of the QFIM are defined as
\begin{equation}\label{qfim}
\mathcal{F}_{\mu\nu}=\mathrm{Re}[\mathrm{Tr}(\rho_{\{\mu\}}L_{\mu}L_{\nu})],
\end{equation}
where the Hermitian operator $L_{\mu}$ is the symmetric logarithmic derivative (SLD) with respect to the parameter $\mu$, defined implicitly by
$\partial\rho_{\{\mu\}}/\partial\mu=\frac{1}{2}\left(\rho_{\{\mu\}}L_{\mu}+L_{\mu}\rho_{\{\mu\}}\right)$.
Equation~(\ref{QCRB}) is in general a matrix inequality, but in the special case of single-parameter estimation it reduces to the scalar Cram\'{e}r-Rao inequality $\delta^2 \mu\geq \mathcal{F}_{\mu\mu}^{-1}$, where $\delta^2 \mu$ is the variance of the  estimator of the parameter $\mu$, and $\mathcal{F}_{\mu\mu}$ is the corresponding quantum Fisher information (QFI).
The variance $\delta^2 \mu$  can be minimized by selecting the probe state with maximal sensitivity to changes in the parameter $\mu$ and by performing the optimal measurement which is given by the projectors onto the eigenvectors of $L_{\mu}$. This leads asymptotically to the saturation of the scalar Cram\'{e}r-Rao bound for any single-parameter estimation.

In multiparameter estimation protocols there are more challenges.
Here, in principle one would like to estimate each parameter as precisely as when using the optimal scheme for estimating that parameter alone, assuming that the other parameters are perfectly known. This is possible when the parameters are \emph{compatible}, that is, they satisfy the following conditions \cite{Ragy}:
(i) There is a single probe state yielding the optimal QFI for each of the parameters.
(ii) There is a single measurement which is optimal for extracting information on all parameters from the evolved state, ensuring the saturability of the quantum Cram\'{e}r-Rao bound; this holds iff
\begin{equation}\label{compa2}\mathrm{Im}[\mathrm{Tr}(\rho_{\{\mu\}}L_{\mu}L_{\nu})]=0 \quad \forall\ \mu, \nu.
\end{equation}
Finally,
(iii) the parameters should be statistically independent, meaning that the indeterminacy of one of them does not affect the precision of estimating the others, which holds only in the case when $\mathcal{F}_{\mu\nu}=0$ for all $\mu\neq\nu$.
A simultaneous estimation scheme requires fewer resources than the corresponding individual estimation scheme by a factor of the number of parameters to be estimated. If the parameters are compatible then no precision is lost for any of the parameters, but the resources required are reduced in this manner so the enhancement attained is maximal.

To get an explicit comparison of performance, we consider the ratio between the minimal total variance of estimating the parameters in the individual and simultaneous schemes as
\begin{equation}
\label{ratio}
R=\frac{\Delta_\mathrm{ind}}{\Delta_{\mathrm{sim}}},
\end{equation}
where
$\Delta_{\mathrm{ind}}=\sum_{\mu}\delta^2 \mu_\mathrm{ind}=\sum_{\mu}\mathcal{F}_{\mu\mu}^{-1}$ and
$\Delta _{\mathrm{sim}}=\frac{1}{p}\mathrm{Tr}(\mathcal{F}^{-1})$,
where $p$ is the number of parameters to be estimated, a factor required to account for the reduction in resources mentioned above.
Hence, $R \leq p$ in general, and $R>1$ indicates that estimating the parameters simultaneously provides an advantage over the individual estimation scheme.

In order to assess the performance of a metrological strategy, it is necessary to define the resources utilized \cite{Ragy}.
To this end, we consider the number of channel applications, $N$, as a common and fixed resource for each strategy.
This choice allows us to interpolate between sequential and parallel strategies to determine the configuration of the optimal protocol, as shown in Figure~\ref{interpolate}. This also allows one to obtain a clear outlook on the role of entanglement in parameter estimation by comparing the sensitivity of entangled states of various size. We will also compare the quantum strategies to the corresponding classical (SQL) strategy by investigating the use of a pure product state as the probe state.


\section{The noise model}\label{sec:Noise}

For the noiseless transformation on our system, let a parameter $\varphi$ be unitarily encoded on a qubit probe by a phase shift $U_{\bm{n}}=e^{-i\varphi H_{\bm{n}}}$ (here using natural units, $\hbar=1$) around the axis $\bm{n}=(\cos\phi \sin\theta,\sin\phi \sin\theta,\cos\theta)$ with $\theta$ and $\phi$ referring to the polar and azimuthal angles on the Bloch sphere.
The generator for this shift is given by $H_{\bm{n}}=\bm{n}\cdot\bm{\sigma}$, with $\bm{\sigma}$ as the vector of the three Pauli matrices.
For $2$-dimensional evolved state $\rho_\varphi=U_{\bm{n}}\rho_{0}U_{\bm{n}}^{\dagger}$, the QFI can be expressed by
$\mathcal{F}_\varphi=|\bm{r}_{0}|^{2}(1-(\bm{r}_{e}\cdot\bm{n}))$ where $\bm{r}_{0}$ and $\bm{r}_{e}$ are the Bloch vectors of the initial state and any eigenstate of it, respectively \cite{Liu}.
It can be found that the maximum value of this QFI is $\mathcal{F}_\varphi^{max}=1$ which can be saturated when $|\bm{r}_{0}|=1$ and $\bm{r}_{0}\perp\bm{n}$.

\begin{figure}[t]
\includegraphics[width=6.5cm]{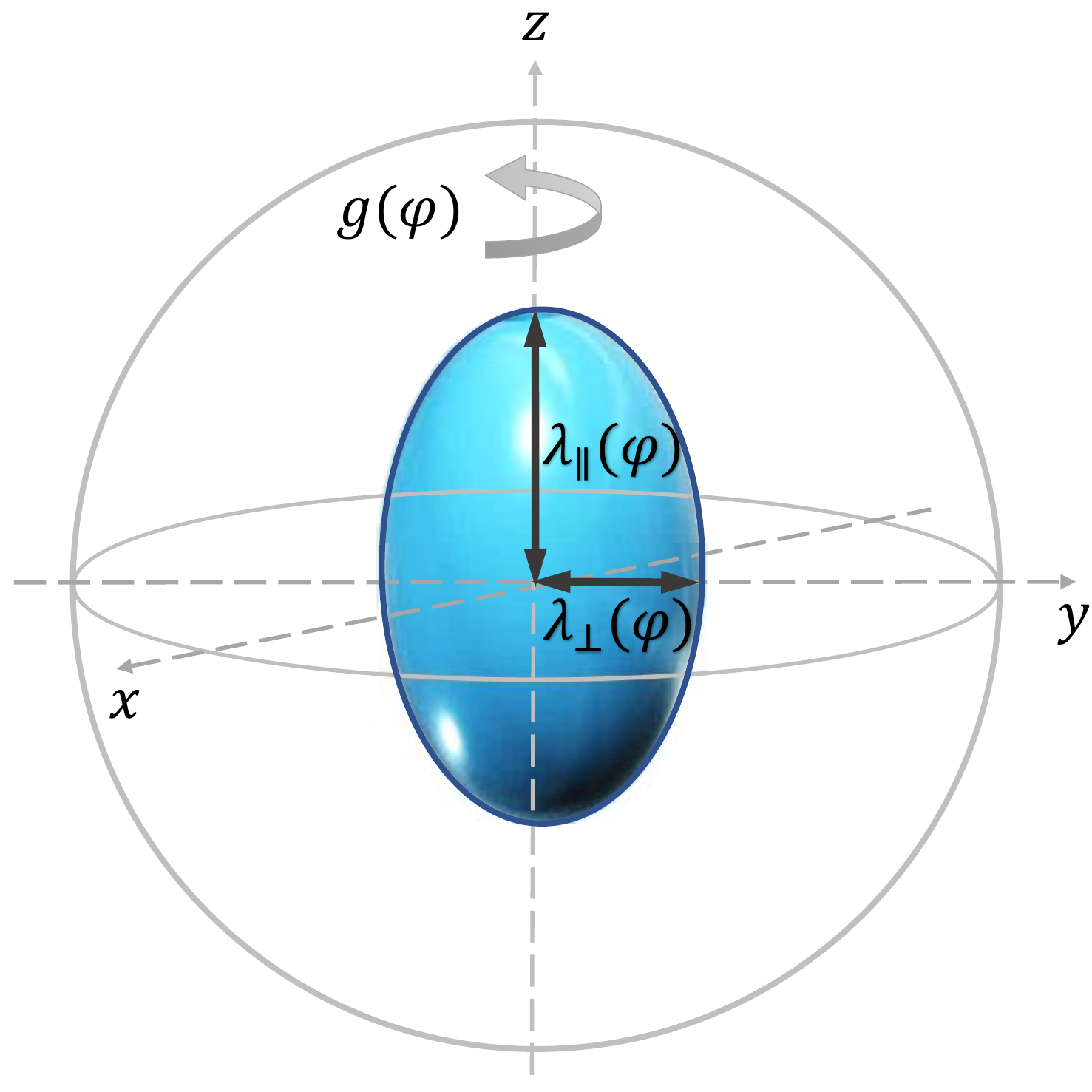}
\caption{Bloch ball representation of a qubit noisy evolution, modelled by a phase-covariant unital channel, whose action consists of: a rotation around the $z$ axis by an angle $g(\varphi)$, a contraction in the horizontal plane by factor $ \lambda_{\perp}(\varphi)$, as well as a contraction in the $z$ direction by factor $\lambda_{\parallel}(\varphi)$. All the plotted quantities are dimensionless.}
\label{fig1}
\end{figure}

In our model, the interaction of the system with its environment results in fluctuations of the transformation such that the direction of the generator $\bm{n}$ is randomly sampled from some normalized probability distribution $p(\theta,\phi)$. Given that no information is available to the experimenter about the specific setting of $\bm{n}$ in each run (in contrast to the studies in \cite{interpower}), the output state after the shift needs to be evaluated by averaging over the prior distribution $p(\theta,\phi)$, which induces an effective noise. Thus the phase imprinting operation alters the probe state $\rho_0$ as \cite{Rosanna}
\begin{equation}
\rho_0 \longrightarrow \Lambda[\rho_0] \equiv \int_0^\pi\!\! d\theta\int_0^{2\pi}\!\! d\phi\, U_\mathbf{n}\,\rho_0\,U_{\mathbf{n}}^\dagger\, p(\theta,\phi)\,\sin{\theta}.
\label{eq:channel}
\end{equation}
The resulting qubit channel $\Lambda$, which encodes both the unitary effect of the phase shift and the noise due to the randomness of its generator, is CPTP and unital (i.e.~$\Lambda[\mathbb{I}]=\mathbb{I}$).
By considering $\langle\bm{n}\rangle=\int_{0}^{2\pi}d\varphi\int_{0}^{\pi}d\theta\, \bm{n}\, p(\theta,\varphi) \sin\theta $ proportional to $(0,0,1)$, that is, fluctuations of the generator direction around the Bloch $z$ axis, one naturally restricts to probability distributions with axial symmetry on the Bloch sphere. This restriction results in this noise model belonging to the physically relevant class of unital phase-covariant qubit channels \cite{covariant channel}.
The overall effect of such noise is to shrink the Bloch ball by factors $\lambda_{\parallel}(\varphi)$ and $\lambda_{\perp}(\varphi)$ in the vertical direction and the
horizontal plane, respectively  (see Figure~\ref{fig1}). It also rotates the ball by an angle $g(\varphi)$, which is not necessarily equal to $\varphi$ as it would be in the noiseless case.
It is important to note that these channels encode information about phase not only in the rotation by $g(\varphi)$, but also in its deformation, such that the information is a function of the noise parameters $\lambda_{\parallel}(\varphi)$ and $\lambda_{\perp}(\varphi)$.

In particular, following \cite{Rosanna} we choose the von Mises-Fisher (vMF) distribution \cite{von Mises-Fisher} with concentration parameter $\kappa$ as the probability distribution for the generator. This has properties analogous to a Gaussian over a sphere and arises naturally for data distributed over such a space. The vMF distribution is defined as
\begin{equation}\label{von Mises Fisher distribution}
p_{\kappa}(\theta)=\frac{\kappa e^{\kappa \cos(\theta)}}{4\pi \sinh(\kappa)}
\end{equation}
Its concentration parameter $\kappa$ gives an idea of how clustered the distribution of $\bm{n}$ is around the mean direction (here, the $z$ axis).
For $\kappa\rightarrow 0$ the probability distribution becomes uniform on the Bloch sphere. Therefore, low $\kappa$ corresponds to a very broad distribution for the random generator direction and, hence, strong noise.
By increasing $\kappa$, the probability distribution becomes more concentrated around the $z$ axis which decreases the strength of the noise such that for $\kappa\rightarrow \infty$, the noise vanishes ($\lambda_{\parallel}=\lambda_{\perp}=1$).

The model when using the vMF distribution may be stated in Liouville form, that is, a matrix representation of the map $\Lambda$ that acts on the four-component vector $|\rho_{0})=(\langle0|\rho_{0}|0\rangle,\langle0|\rho_{0}|1\rangle,\langle1|\rho_{0}|0\rangle,\langle1|\rho_{0}|1\rangle)^{T}$, as
\begin{equation}\label{Map's matrix}
\tilde{\Lambda} = \begin{pmatrix} \frac{1+\lambda_{\parallel}(\varphi)}{2} & 0 & 0 &\frac{1-\lambda_{\parallel}(\varphi)}{2}\\ 0 & \lambda_{\perp}(\varphi)e^{-ig(\varphi)} & 0 & 0\\ 0 & 0 & \lambda_{\perp}(\varphi)e^{ig(\varphi)} & 0 \\ \frac{1-\lambda_{\parallel}(\varphi)}{2} & 0 & 0 & \frac{1+\lambda_{\parallel}(\varphi)}{2} \end{pmatrix},
\end{equation}
where
\begin{eqnarray}
\nonumber \lambda_{\parallel}(\varphi) = 1-2b, &&\quad  \lambda_{\perp}(\varphi)=|c|,\\
\nonumber \cos g(\varphi)&=& \frac{c+c^{\ast}}{2|c|},
\end{eqnarray}
and
\begin{eqnarray}\label{bc}
\nonumber b &=& \frac{2\sin^{2} \varphi}{\kappa^{2}}(\kappa \coth \kappa -1), \\
  c &=& \cos 2\varphi +b(1-i \kappa \cot \varphi)
\end{eqnarray}

In \cite{Rosanna}, estimation of the single parameter $\varphi$ under this noise model was analyzed, assuming the specifics of the probability distribution, and in particular the concentration $\kappa$, known a priori. In this work, we consider instead the more realistic situation in which the concentration (which determines the strength of the noise as discussed above) is itself a parameter to be estimated, and investigate whether an enhancement may be attained by simultaneously measuring $\varphi$ and $\kappa$ in a multiparameter estimation scheme.


\section{Multiparameter estimation with single qubit probes}\label{sec:1qubit}

Studying the precision attainable with single qubit probe states is not only instructive in revealing the physics behind metrology at this scale, but is also necessary to analyze the precision in classical estimation schemes.
This is because the QFI of $M$ identical uncorrelated subsystems is just $M$ times the QFI of each single one of such subsystems, which follows from the convexity of the QFI on density matrices \cite{Convexity}, and leads to the scaling of the precision in a classical strategy being governed by the SQL.

Taking a single qubit probe system in the initial state $\rho_{0}$, with $\bm{r}_{0}=(\langle \sigma_{x}\rangle_{\rho_{0}},\langle \sigma_{y}\rangle_{\rho_{0}},\langle \sigma_{z}\rangle_{\rho_{0}})^{T}$ as its corresponding Bloch vector, one can obtain
the final state $\rho_{\varphi,\kappa}=\Lambda[\rho_{0}]$ with a mapped Bloch vector $\bm{r}=(r_{1},r_{2},r_{3})^{T}$ given by
\begin{eqnarray}
\nonumber  r_{1} &=& (\cos(2\varphi)+b)\langle \sigma_{x}\rangle_{\rho_{0}}-b \kappa \cot(\varphi)\langle \sigma_{y}\rangle_{\rho_{0}} \\
\nonumber  r_{2} &=& (\cos(2\varphi)+b)\langle \sigma_{y}\rangle_{\rho_{0}}+b \kappa \cot(\varphi)\langle \sigma_{x}\rangle_{\rho_{0}} \\
\nonumber  r_{3} &=& (1-2b)\langle \sigma_{z}\rangle_{\rho_{0}}.
\end{eqnarray}
where $b$ is given in Eq.~\eqref{bc}. Since $|\bm{r}|< 1$, an exponential form can be obtained for this state, $\rho_{\varphi,\kappa}=e^{G}$ with
\begin{equation}\label{G}
G=\frac{1}{2}\ln\left(\frac{1-|\bm{r}|^{2}}{4}\right)\mathbb{I}+\frac{\tan^{-1}\left(|\bm{r}|\right)}{|\bm{r}|}\bm{r}\cdot\bm{\sigma}
\end{equation}
Following the working of \cite{QFI for states in exponential form}, the SLD may be written as
\begin{equation}\label{SLD}
L_{\mu}=\frac{\bm{r}\cdot(\partial_{\mu}\bm{r})}{1-|\bm{r}|^{2}}(-\mathbb{I}+\bm{r}\cdot\bm{\sigma})+(\partial_{\mu}\bm{r})\cdot\bm{\sigma},
\end{equation}
which gives the entries of the QFIM as
\begin{equation}\label{Fisher}
\mathcal{F}_{\mu\nu}=(\partial_{\mu}\bm{r})\cdot(\partial_{\nu}\bm{r})+\frac{(\bm{r}\cdot\partial_{\mu}\bm{r})(\bm{r}\cdot\partial_{\nu}\bm{r})}{1-|\bm{r}|^{2}}
\end{equation}
for $\mu,\nu\in\{\varphi,\kappa\}$.

\begin{figure}[t]
\centering
\includegraphics[width=7.5cm]{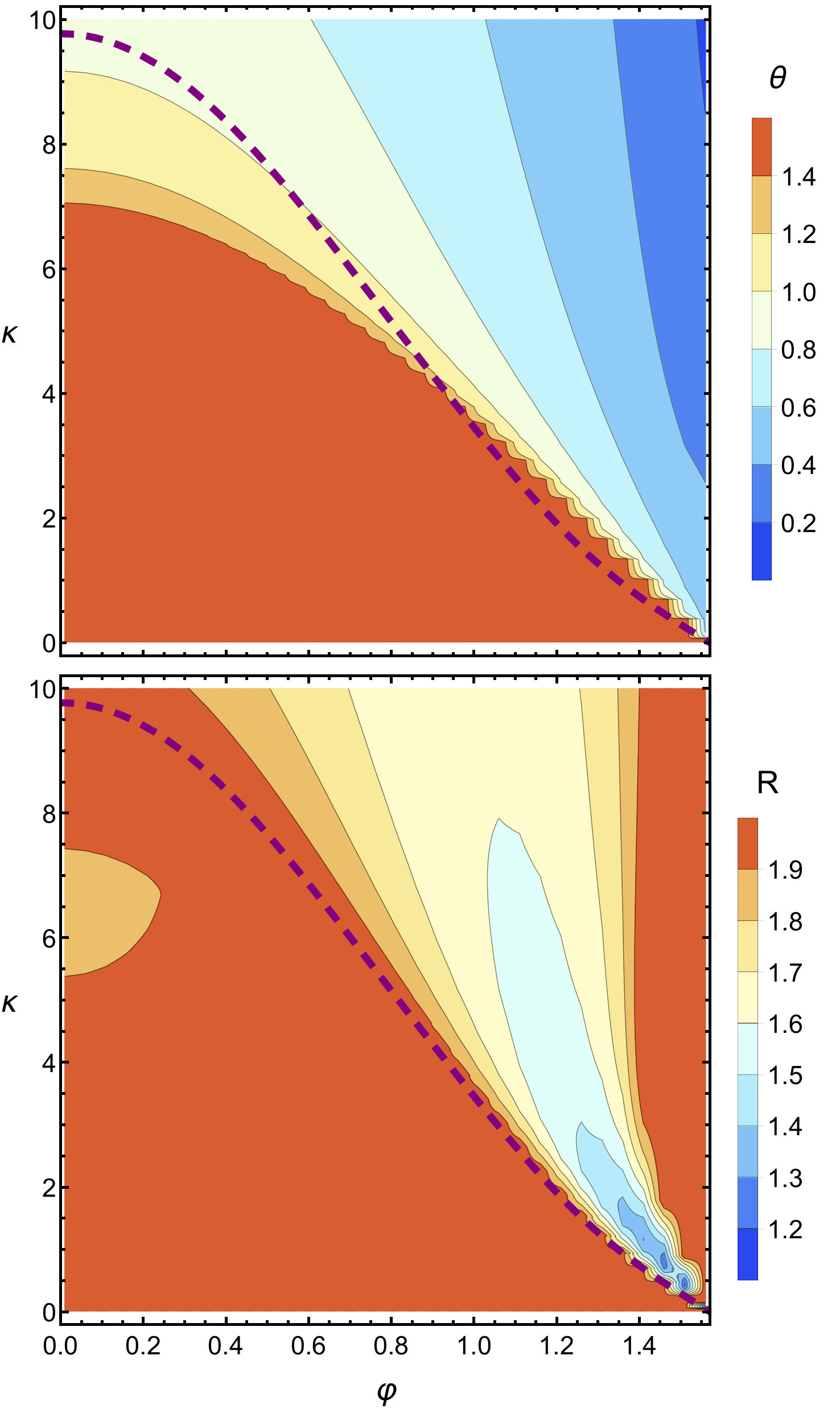}
\caption{Multiparameter estimation with single qubit probes. Top: The variation of the optimal state for simultaneous estimation with the parameters $\varphi$ and $\kappa$.  The purple dashed line separates the two possible optimal initial states to estimate $\kappa$ individually: below the line, any equatorial state, characterized by $\theta = \pi/2$, is optimal, while above the optimal state becomes the one with $\theta = 0$.  The equatorial state is always optimal when estimating $\varphi$ individually instead. Note moreover that the regions where the equatorial state is optimal for the individual estimation of $\kappa$ and for the joint estimation of both parameters simultaneously are not the same. Bottom: Ratio $R=\Delta_{\mathrm{ind}}/\Delta_{\mathrm{sim}}$ against the parameters to be estimated. This shows $R>1$, and thus the superiority of the simultaneous scheme, over all parameter space. All the plotted quantities are dimensionless.}
\label{fig:onequbit}
\end{figure}

To determine whether the parameters $\varphi$ and $\kappa$ are compatible in a multiparameter estimation scheme, the conditions (i), (ii) and (iii), as outlined in Section \ref{sec:QuantumParameterEst}, need to be addressed.
The first requires that both parameters may be optimally estimated using the same initial probe state. When estimating $\varphi$ individually, the QFI obeys,
\begin{equation}
\mathcal{F}_{\varphi\varphi}\leq|\partial_{\varphi}c|^{2}+\frac{(\partial_{\varphi}|c|)^{2}}{1-|c|^{2}}\equiv\mathcal{F}_{\varphi\varphi}^{\mathrm{max}},
\end{equation}
where the inequality can be saturated for any state on the equator of the Bloch sphere, i.e.~$\theta = \pi/2$.
An analogous condition cannot be obtained for the maximum QFI when individually estimating the noise parameter $\kappa$.
The corresponding quantity, $\mathcal{F}_{\kappa\kappa}^{\mathrm{max}}$, has two possible forms, depending on the values of the parameters. For low $\varphi$ and $\kappa$ it is given by $|\partial_{\kappa}c|^{2}+\frac{(\partial_{\kappa}|c|)^{2}}{1-|c|^{2}}$, which may be obtained by using any equatorial state. At high $\varphi$ and $\kappa$, this becomes $\frac{(\partial_{\kappa}b)^{2}}{b(1-b)}$ which is obtained when $\theta = 0$, i.e., for a state at the north pole. Therefore, the first compatibility condition is only satisfied in the low parameter regime. This transition is shown in Figure~\ref{fig:onequbit}.

For sufficiently low values of the parameters $\varphi$ and $\kappa$, the optimal probe state for the simultaneous scheme is still any equatorial state. Curiously, this region in parameter space is not precisely the same as that where these states are also optimal for estimating $\kappa$ individually, but lies inside the latter. Beyond this, the optimal state rises above the equator, approaching the north pole as the parameters increase (see Figure~\ref{fig:onequbit}).

The equality
\begin{equation}\label{necessary condition1}
\mathrm{Tr}(\rho_{\varphi,\kappa}[L_{\varphi},L_{\kappa}])=2i \bm{r}\cdot (\partial_{\varphi}\bm{r}\times\partial_{\kappa}\bm{r})=0,
\end{equation}
may only be satisfied with the equatorial or polar state, but not with any superposition of the two. Therefore, condition (ii) is also only satisfied in the low parameter regime. We must therefore be wary that over the transition, the quantum Cram\'{e}r-Rao bound bound may not be saturated in the simultaneous scheme.

Compatibility condition (iii) requires instead the vanishing of the off-diagonal elements of the QFIM, which is not achieved in this scheme.

This clearly shows that the multiparameter scheme with one qubit cannot fully meet the compatibility conditions.
Nonetheless, an advantage may still be attained by simultaneously estimating the parameters.
To compare the two strategies we must first define the resources precisely. For these schemes, we are in fact comparing the performance of two uncorrelated qubit probes. In the independent scheme one of these is used to estimate $\kappa$ and the other is used to estimate $\varphi$, thus giving the minimal total variance on both the parameters as $\Delta_{\mathrm{ind}}(2)=\sum_{\mu}\delta^2 \mu_{\mathrm{ind}}(1)=\mathcal{F}_{\varphi\varphi}(1)^{-1}+\mathcal{F}_{\kappa\kappa}(1)^{-1}$. In simultaneous estimation each single qubit is used to estimate both parameters, so the results from the two qubits may be combined classically to give $\Delta _{\mathrm{sim}}(2)=\frac{1}{2}\Delta _{\mathrm{sim}}(1)=\frac{1}{2}\mathrm{Tr}\left[\mathcal{F}^{-1}(1)\right]$. When the parameters are compatible, the off-diagonal elements of the  QFIM are zero, giving $\Delta _{\mathrm{sim}}(2)=\frac{1}{2}\Delta_{\mathrm{ind}}(2)$. The ratio $R=\Delta_{\mathrm{ind}}/\Delta_{\mathrm{sim}}$ introduced in Eq.~(\ref{ratio}) therefore has a maximum of 2, and $R > 1$ indicates superiority of the simultaneous scheme.
In Figure~\ref{fig:onequbit} we see that the simultaneous strategy is always advantageous and is very close to the ideal, compatible, case ($R=2$) in the low parameter region.

\section{Multiparameter estimation with bipartite probes}\label{sec:2qubit}

In the context of phase estimation, the role of entanglement in the preparation stage has attracted remarkable attention in recent years \cite{Giovannetti,Braun,Pezze}.
To begin with the exploration of its role in our multiparameter estimation problem, let us first treat the simplest case of two qubit entangled probes.

Due to the axial symmetry of the model described in Section~\ref{sec:Noise}, the probe states that allow optimal parameter estimation can be sought in the class of states exhibiting both permutational symmetry of the qubits, and parity symmetry under bit flips.
We thus consider the following class of two-qubit states
\begin{equation}\label{initial two-qubit}
|\psi\rangle=\alpha(|00\rangle+|11\rangle)+\beta(|01\rangle+|10\rangle),
\end{equation}
with $\alpha,\beta\in\mathbb{R}$ and $\alpha^{2}+\beta^{2}=\frac{1}{2}$.
When using the two qubits in parallel, the channel $\Lambda$ acts independently on each qubit, yielding
\begin{eqnarray}
\nonumber  \rho_{\varphi,\kappa} = \Lambda\otimes\Lambda[|\psi\rangle\langle\psi|]
   = \begin{pmatrix} \alpha^{2}-\xi & \alpha\beta c^{\ast} & \alpha\beta c^{\ast} & \alpha^{2} c^{\ast2}\\ \alpha\beta c & \beta^{2}-\xi & \beta^{2}|c|^{2} & \alpha\beta c^{\ast}\\ \alpha\beta c & \beta^{2}|c|^{2} & \beta^{2}-\xi & \alpha\beta c^{\ast} \\ \alpha^{2} c^{2} & \alpha\beta c & \alpha\beta c & \alpha^{2}-\xi \end{pmatrix},\\
\end{eqnarray}
where $\xi=2b(1-b)(\alpha^{2}-\beta^{2})$.

\begin{figure}
    \centering
    {
        \includegraphics[width=4.2cm]{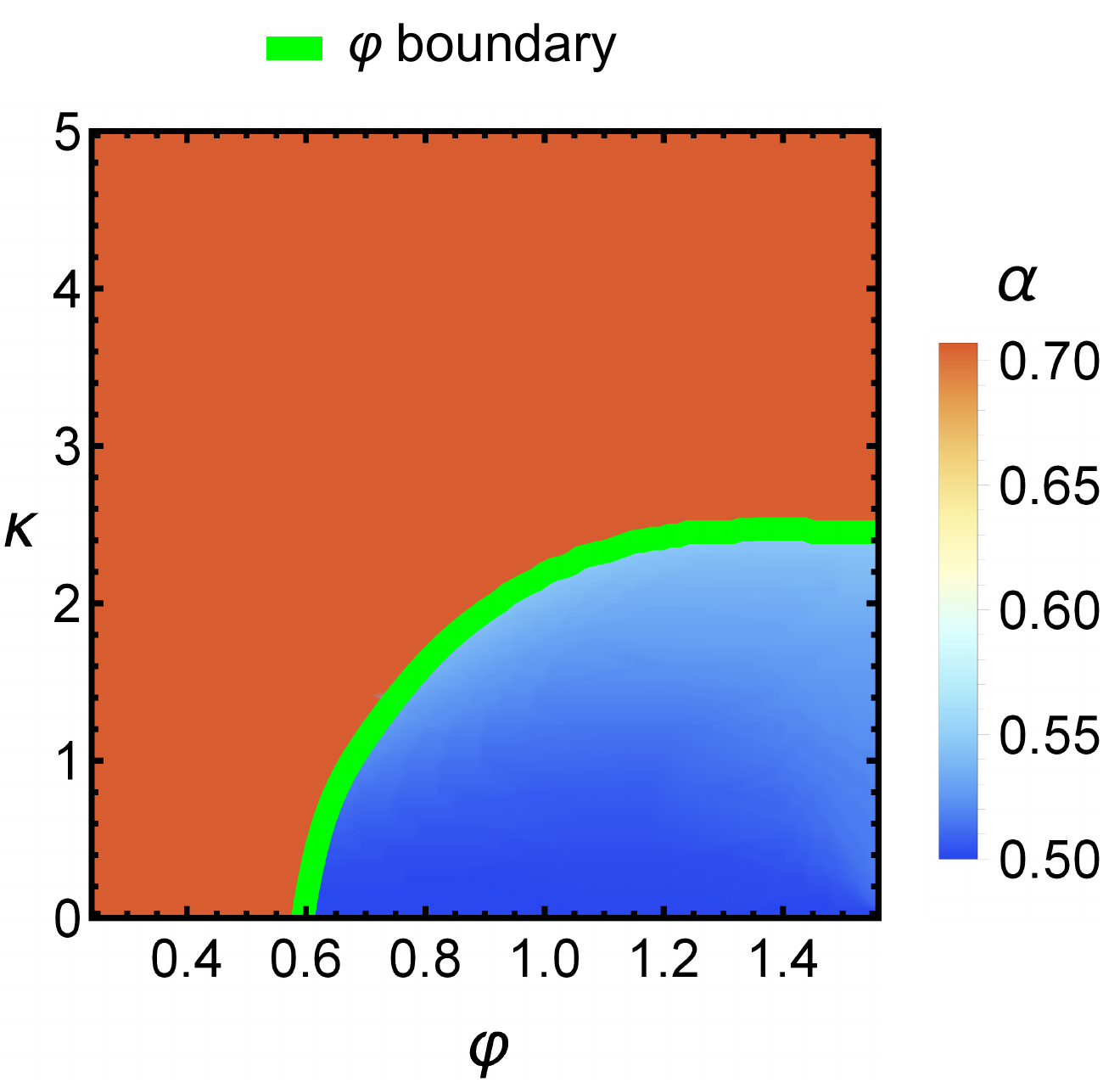}
    }
    {
        \includegraphics[width=4.2cm]{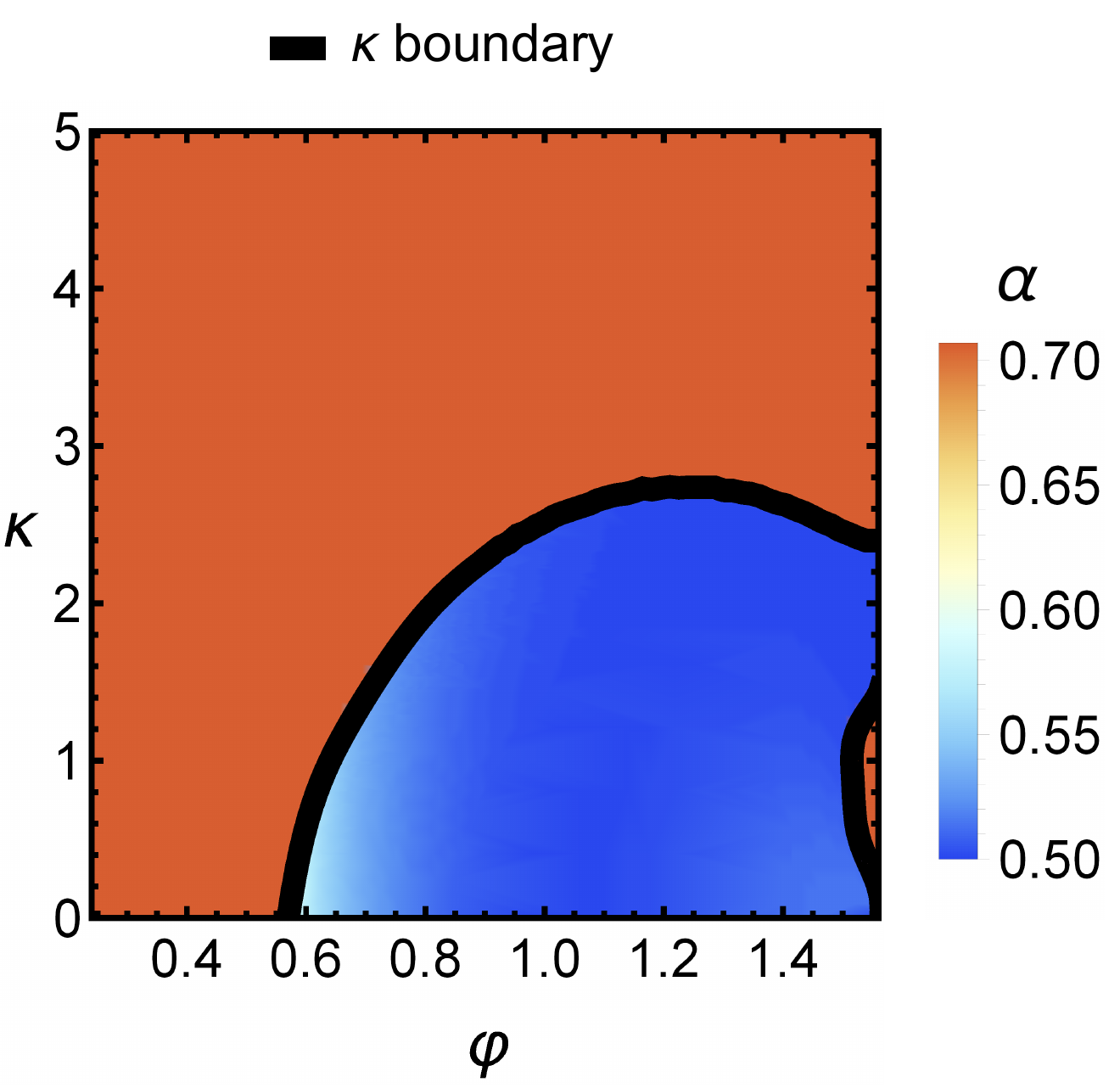}
    }
    \caption{(color online) Optimal bipartite probe states to individually estimate (Left): $\varphi$ and (Right): $\kappa$, as specified by the parameter $\alpha$ in the two qubit state $\ket{\psi}$ of Eq.~(\ref{initial two-qubit}), with $\beta = \sqrt{\frac{1}{2}-\alpha^2}$.  The thick boundaries delimit the region where the maximally entangled state ($\alpha=1/\sqrt{2}$) ceases to be optimal and is eventually superseded by a product state $(\alpha=1/2)$. The corresponding plot in the case of simultaneously estimating both parameters is almost indistinguishable from the one of estimating $\kappa$ alone (Right), and is not shown here. All the plotted quantities are dimensionless.}
    \label{twoqubit}
\end{figure}

After a tedious but straightforward calculation, we find that when estimating either parameter individually, or when estimating them both simultaneously, the optimal probe state is given by the maximally entangled state  $\frac{1}{\sqrt{2}}(|00\rangle+|11\rangle)$
for most of the relevant parameter space. However, for the strong noise regime of high $\varphi$ and low $\kappa$, the optimal state becomes the product state $|+\rangle^{\otimes 2}$ with $|+\rangle=\frac{1}{\sqrt{2}}\left(|0\rangle+|1\rangle\right)$.
This is the case when the quantum estimation protocol, relying on entanglement, is no longer advantageous but is outperformed by the classical strategy relying on local coherence. This quantum-classical boundary is slightly different for each scheme (independent versus simultaneous) and in both cases is slightly blurred, such that for some parameter values, the optimal $\alpha$ lies between $1/2$ and $1/\sqrt{2}$. This is shown in Figure~\ref{twoqubit}.

Note that any probe state described by Eq.~(\ref{initial two-qubit}) satisfies condition (ii) of the compatibility requirements, given by Eq.~(\ref{compa2}), implying that the quantum Cram\'{e}r-Rao bound can always be saturated in this scheme.

We also examine the performance ratio, $R$, when using bipartite probes. Here, the off-diagonal elements of the QFIM do not vanish and so the compatibility condition (iii) is not met, but still the simultaneous scheme may provide some advantage. As before, this advantage is shown when $R>1$.

\begin{figure}
    \centering
    {
        \includegraphics[width=7.5cm]{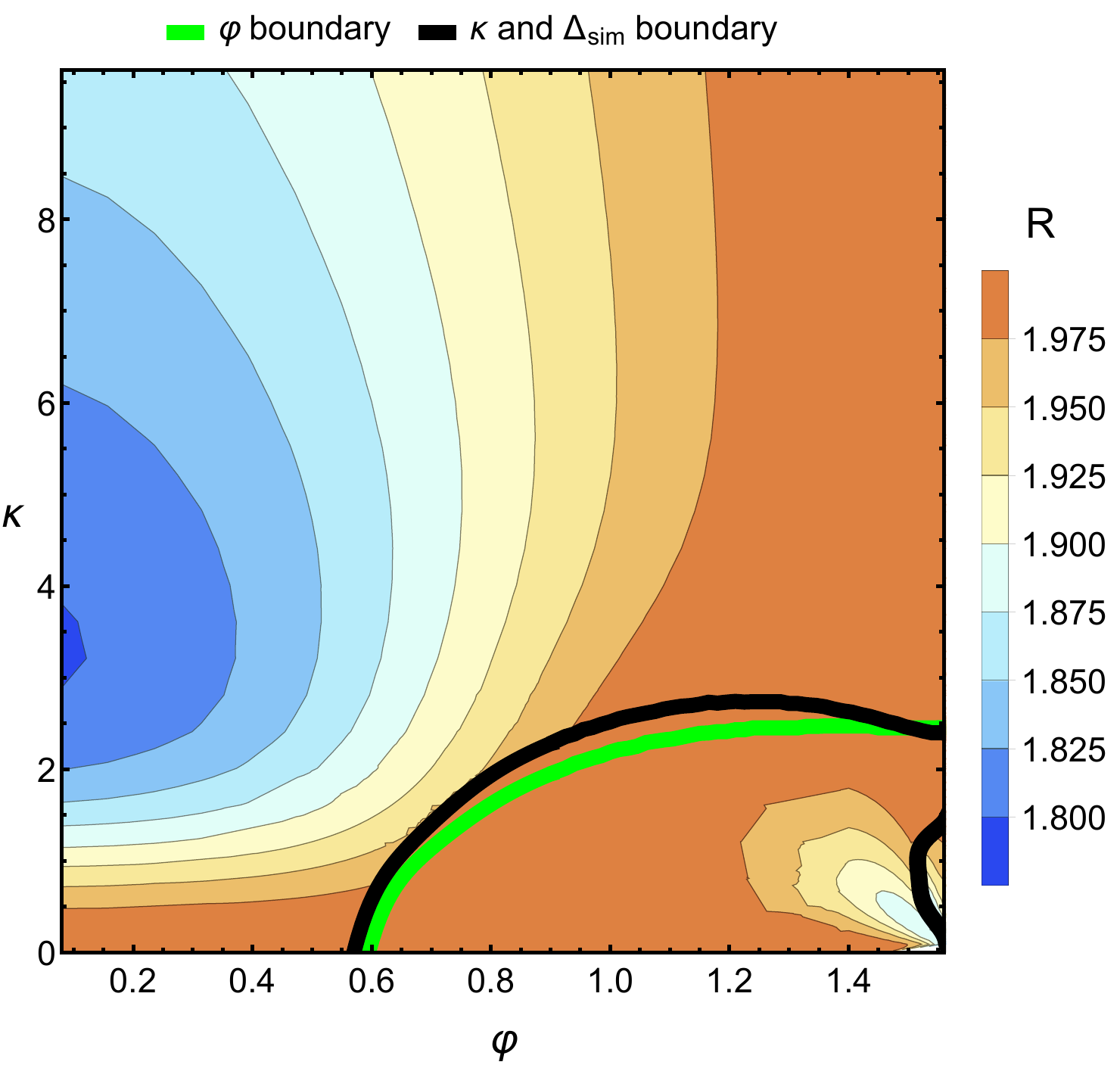}
    }
    \caption{The performance ratio  $R=\Delta_{\mathrm{ind}}/\Delta_{\mathrm{sim}}$ versus the parameters to be estimated, using optimal bipartite probe states. Here $R$ is always above 1 and usually close to its maximum, 2, showing the superiority of estimating both parameters simultaneously. The boundaries are the same as in Figure~\ref{twoqubit}. All the plotted quantities are dimensionless.}
    \label{R2}
\end{figure}

As shown in Figure~\ref{R2}, the simultaneous scheme is advantageous for all values of the parameters investigated and is often close to the ideal case, $R=2$,
where the advantage stems from the two-fold reduction in resources required.

\section{Metrological strategies for multiparameter estimation with fully entangled probes}\label{sec:Nqubit}

\begin{figure*}[t]
	\centering
	\includegraphics[height=0.375\linewidth]{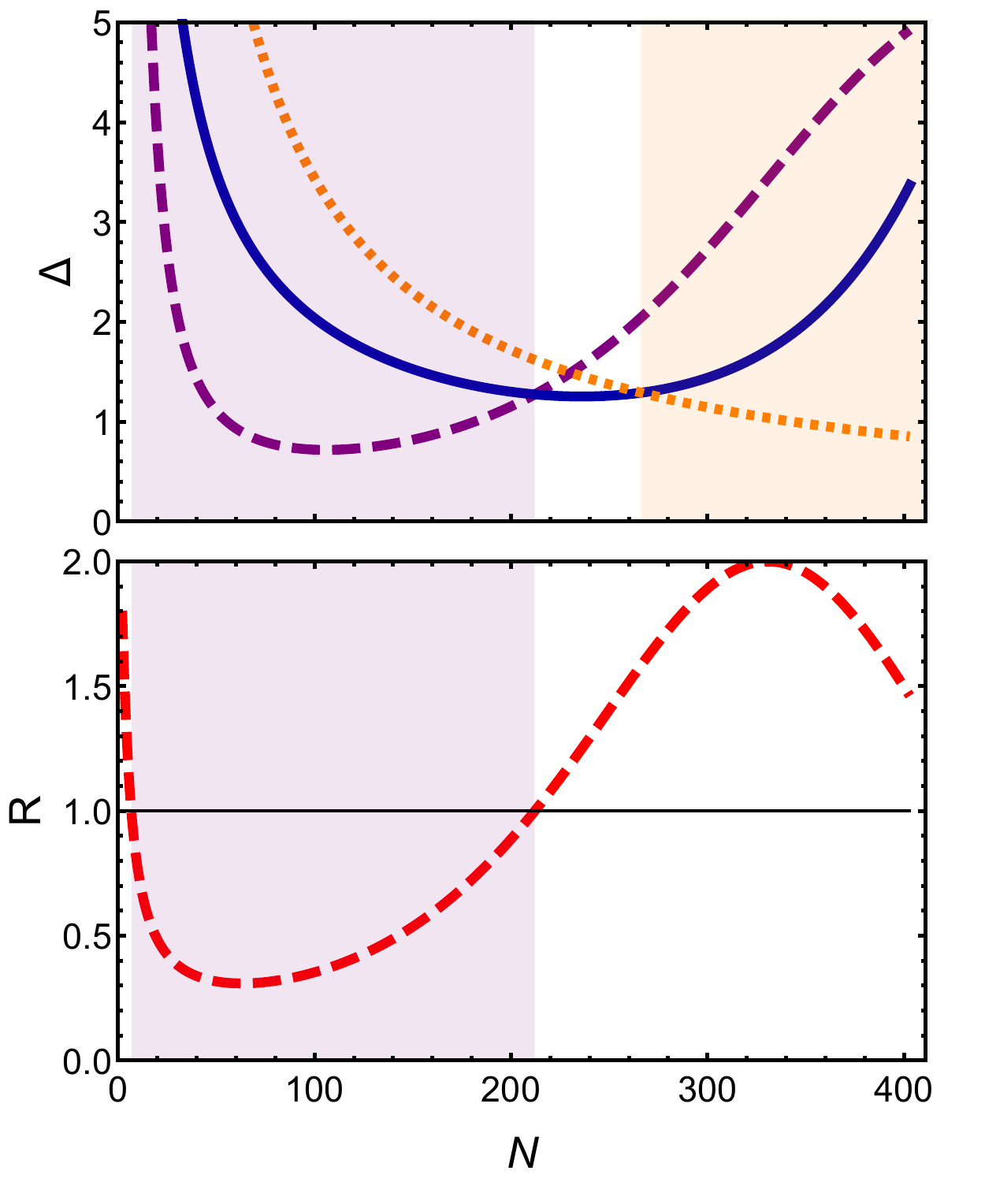}
	\includegraphics[height=0.375\linewidth]{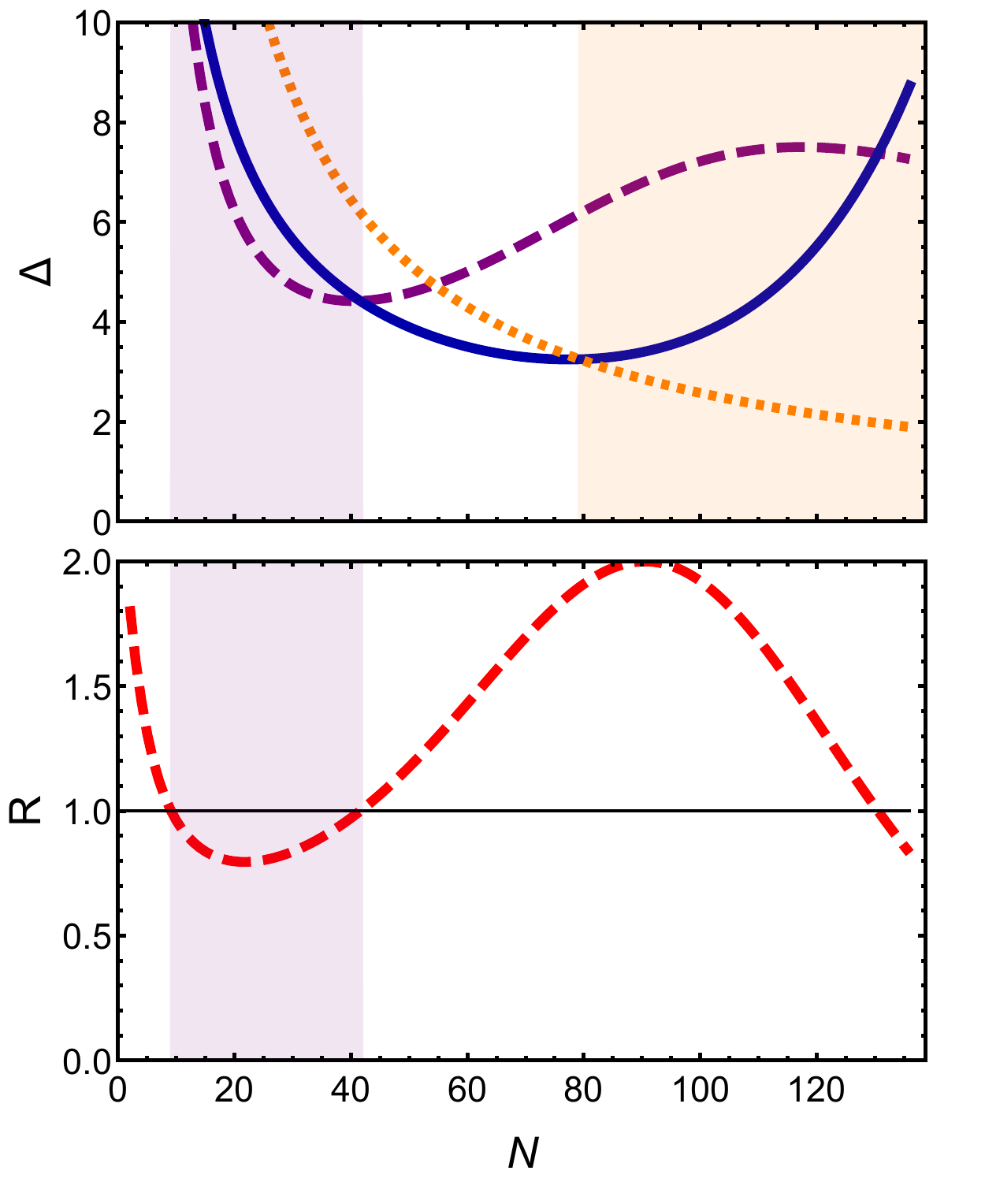}
	\includegraphics[height=0.375\linewidth]{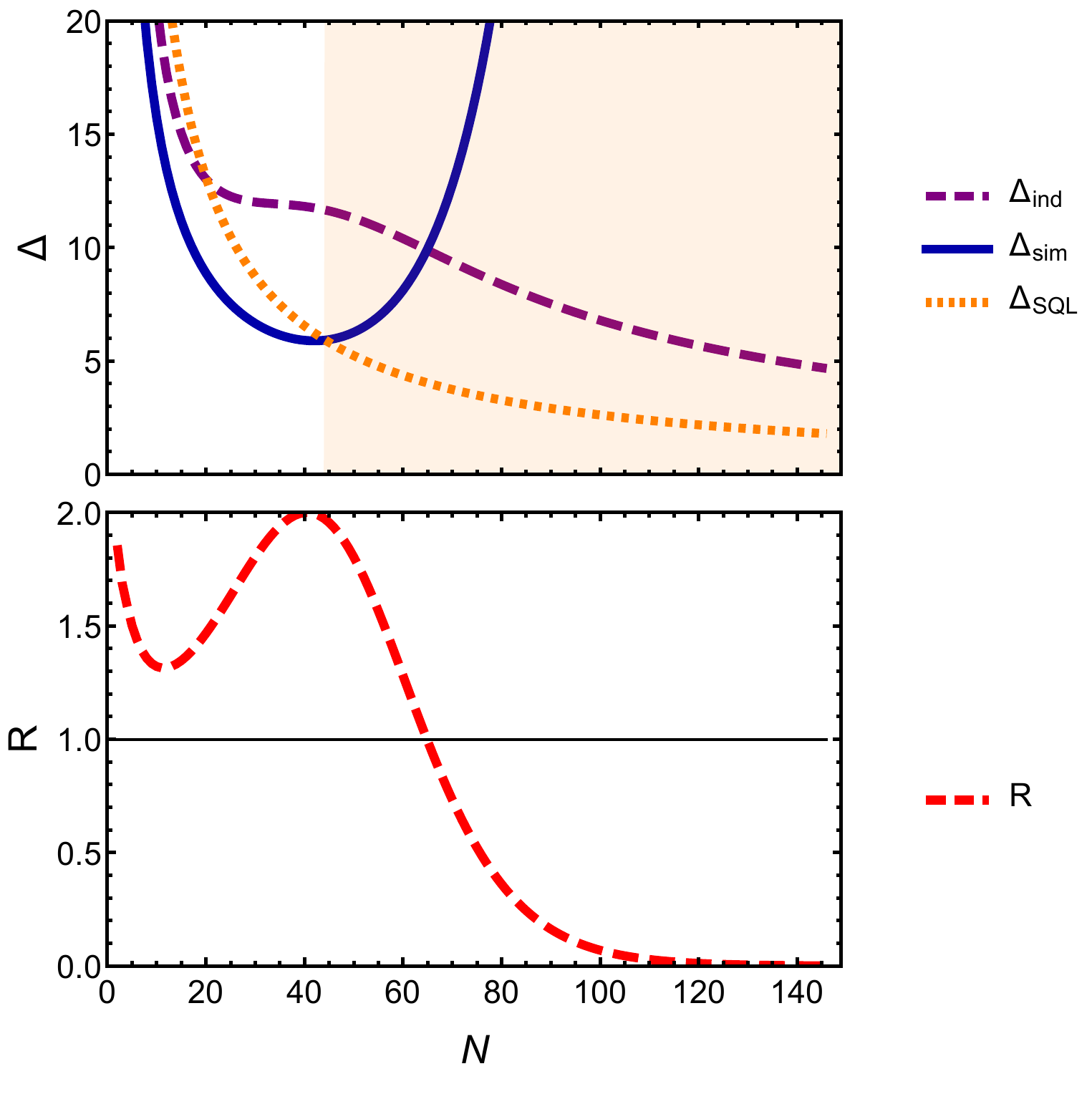}
	\caption{The error associated to each estimation strategy plotted as a function of the number of probe qubits, $N$, showing examples of different metrological regimes. Left: Case A, produced with $\varphi = 0.150$, $\kappa=3.00$. Middle: Case B, produced with $\varphi = 0.312$, $\kappa=4.31$. Right: Case C, produced with $\varphi = 0.500$, $\kappa=5.50$. All the plotted quantities are dimensionless.}
	\label{fig:Cases}
\end{figure*}

We now discuss the metrological strategies available with multipartite entangled probe states. Referring to the scheme of Figure~\ref{interpolate}, here we begin by considering the fully parallel setting, which corresponds to $M=N$. For $N$ qubits, we choose the maximally entangled Greenberger-Horne-Zeilinger (GHZ) state $\frac{1}{\sqrt{2}}\left(\ket{0\ldots 0} + \ket{1 \ldots 1}\right)$ \cite{GHZ} as our probe state. This state is permutationally symmetric and is known to be optimal in noiseless phase estimation \cite{Giovannetti2006}, as well as in the low parameter regime in the two qubit case above for our model, but it may not be so in the presence of noise in general.

Each qubit in the probe undergoes evolution under the action of the channel $\Lambda$ as follows
\begin{eqnarray}\label{twoparts}
\nonumber\rho&=&\Lambda^{\otimes N}[\rho_{0}]\\
& = & \varrho_{1} + \varrho_{2},
\end{eqnarray}
where we make use of the structure of the GHZ state to decompose the evolved state into its corner and diagonal parts
\begin{eqnarray}
\nonumber  \varrho_{1} &=& \frac{1}{2}\left( \left(b^{N}+(1-b)^{N}\right)\left(|0\rangle\langle0|^{\otimes N}+|1\rangle\langle1|^{\otimes N}\right)\right.\\
& &\left.+ c^{N}|0\rangle\langle1|^{\otimes N}+c^{\ast N}|1\rangle\langle0|^{\otimes N} \right),\\
\nonumber \varrho_{2}&=&\frac{1}{2}\sum_{m=1}^{N-1}\left(b^{m}(1-b)^{N-m}+b^{N-m}(1-b)^{m}\right)\\
& & \times\, |0\rangle\langle0|^{\otimes N-m} \,|1\rangle\langle1|^{\otimes m}.
\end{eqnarray}

In our analysis, as usual, one may choose to estimate $\kappa$ and $\varphi$ individually or simultaneously.  Hence, we discuss the following possible strategies using $2N$ qubit probes: the individual estimation, where two $N$-partite GHZ states are used, one to estimate $\kappa$ and one to estimate $\varphi$; the simultaneous estimation, where two $N$-partite GHZ states are used, each estimating the two parameters simultaneously; and the classical (SQL) strategy, where estimates of the parameters are obtained from measurements on $2N$ single qubit states, according to the optimal prescription of Section~\ref{sec:1qubit}. The estimation errors in the individual and simultaneous strategies  may be found from the entries in the QFIM as follows:
\begin{eqnarray}
\label{deltaIndSim}
\nonumber \Delta_\mathrm{ind} (2N) &  =  & \delta^2 \varphi(N) +  \delta^2 \kappa(N) \\
&  =  &  F_{\varphi\varphi}(N)^{-1} + F_{\kappa\kappa}(N)^{-1}, \\
\nonumber \Delta_\mathrm{sim} (2N)& = & \frac{1}{2} \Delta_{\mathrm{sim}}(N) \\
\nonumber & = &  \frac{1}{2} \left(\left(F_{\varphi\varphi}(N)-\frac{F_{\varphi\kappa}(N)^2}{F_{\kappa\kappa}(N)}\right)^{-1} \right.\\
& & \left. + \left(F_{\kappa\kappa}(N)-\frac{F_{\varphi\kappa}(N)^2}{F_{\varphi\varphi}(N)}\right)^{-1}\right).
\end{eqnarray}

For the classical strategy, one must decide whether to use each probe to estimate the parameters simultaneously or whether to estimate the parameters individually, using half the probes for each parameter. In either case, the $1/N$ behavior of the SQL is followed, with the prefactor determined by the error on the single qubit probe estimation strategy.
Therefore, as the simultaneous strategy is always superior at the single qubit level, it will also be optimal for any number of independent qubits, and the error for the classical strategy will be given by
\begin{equation}
\Delta_\mathrm{SQL} (2N) = \frac{1}{2N} \Delta_{\mathrm{sim}}(1)
\end{equation}

Let us now compare the performance of these strategies over a range of the phase parameter, $\varphi$, and noise parameter, $\kappa$, focusing on the low parameter regime (under the dashed line in Figure~\ref{fig:onequbit}). Here, for each given $\varphi$ and $\kappa$ there are three possible behavior regimes as the errors vary with  $N$, with examples of each of these given in Figure~\ref{fig:Cases}. In all three cases there are some common features shown in the estimation strategies. For both quantum estimation strategies, the error initially decreases faster than the SQL, before hitting a minimum and starting to increase once again. The individual estimation strategy reaches a maximum before decaying again, but not faster than the SQL. The simultaneous strategy however does not reach a maximum but continues to increase. As expected, the classical strategy maintains its $1/N$ behavior in all regimes and will always beat the quantum estimation strategies at sufficiently high $N \gg 1$.

\begin{figure*}[t]
	\centering
	\includegraphics[height=7cm]{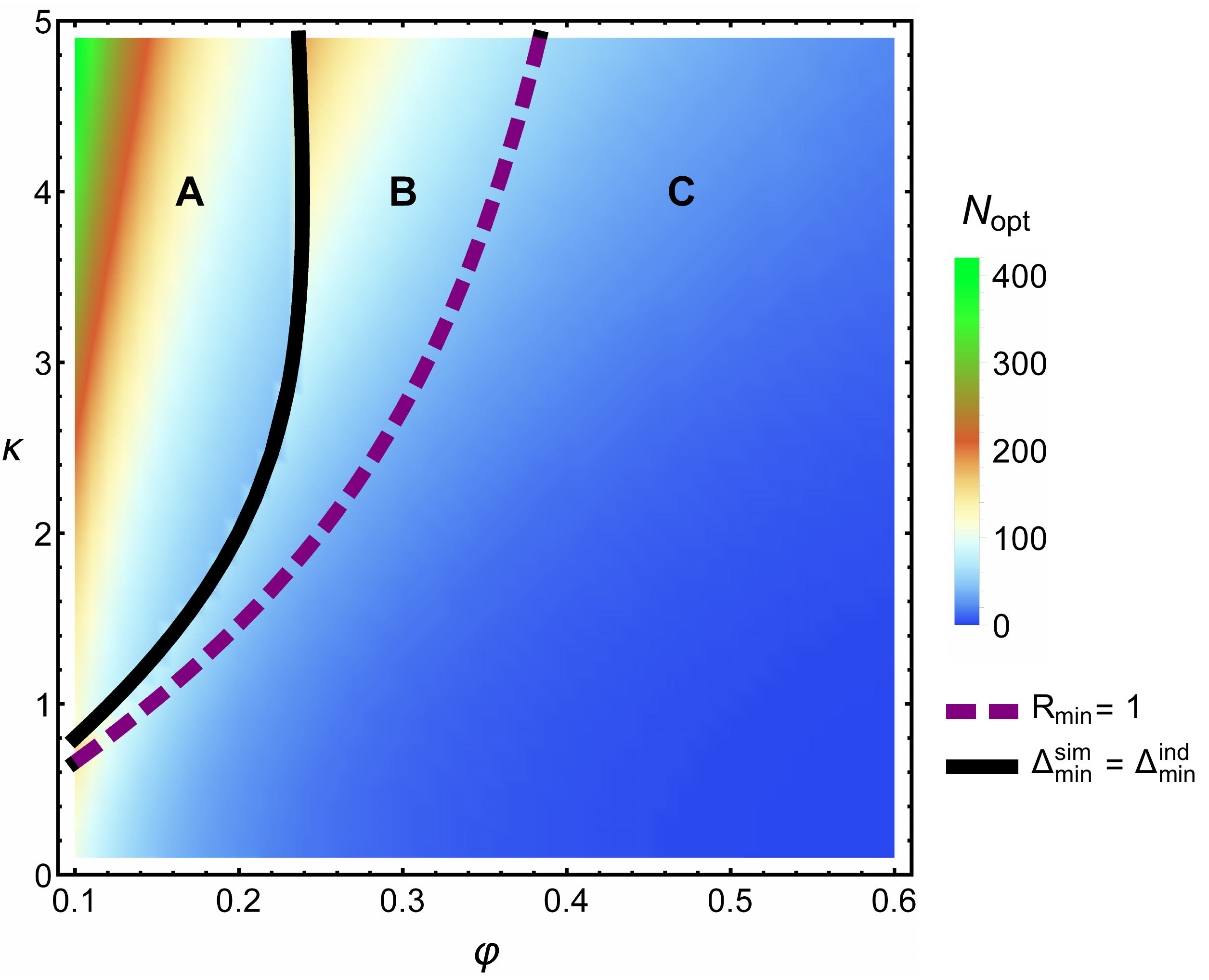} \hspace*{.5cm}
	\includegraphics[height=6.92cm]{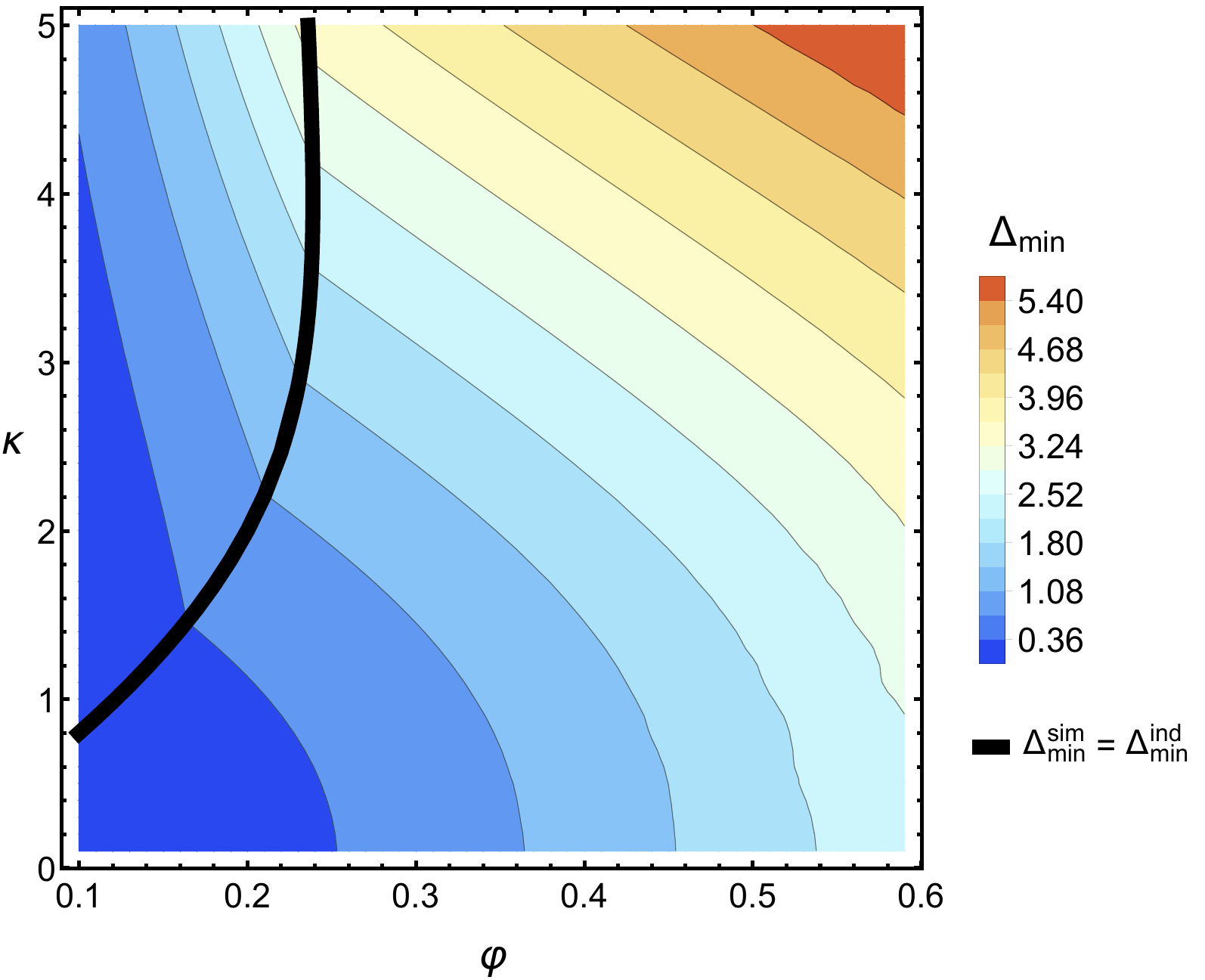} \\
	\caption{Left: The optimal number $N_\mathrm{opt}$ of entangled probes as a function of the phase parameter $\varphi$ and noise parameter $\kappa$. The solid boundary $\Delta_{\mathrm{min}}^{\mathrm{sim}} = \Delta_{\mathrm{min}}^{\mathrm{ind}} $ separates Case A, where the $N_{\mathrm{opt}} = N_{\mathrm{opt}}^{\mathrm{ind}}$, and Case B, where $N_{\mathrm{opt}} = N_{\mathrm{opt}}^{\mathrm{sim}}$. The dashed boundary $R_{\mathrm{min}}=1$ separates Case B, where simultaneous and individual measurements can each give better performance for some $N$, and Case C, where the simultaneous strategy is always preferred, until the classical strategy take over with large $N$. Right: Contour plot of the corresponding optimal error $\Delta_{\min}$ on the estimation of $\varphi$ and $\kappa$ at $N=N_{\mathrm{opt}}$ for the strategies described above. Once more, the solid boundary separates Case A, where the plotted $\Delta_{\min}$ corresponds to the error in the individual strategy, from Cases B and C, where the plotted $\Delta_{\min}$ corresponds to the error in the simultaneous strategy. All the plotted quantities are dimensionless.}
	\label{fig:FullContourPlot}
\end{figure*}

We remark that the compatibility condition (ii) given by Eq.~(\ref{compa2}) is always fulfilled by the GHZ state for any $N$, which ensures that the estimation error arising from the quantum Cram\'er-Rao bound is in principle achievable.
As in previous sections, we use $R$ defined in Eq.~(\ref{ratio}) as an indicator of the best strategy to use, $R$ below 1 indicating the superiority of the individual estimation strategy and $R$ above 1 indicating the superiority of the simultaneous strategy, with a maximum possible value of $R =2$ when all parameter compatibility requirements are fulfilled.
In all regimes, $R$ begins close to 2 (see Figures~\ref{fig:onequbit} and \ref{R2}) before decreasing to a minimum, increasing to 2 and finally decaying to zero at high $N$, due to the asymptotic growth of $\Delta_\mathrm{sim}$ and the decay of $\Delta_\mathrm{ind}$. Whether the initial minimum drops or not below 1 determines which of the three behavior cases the regime falls under, as distinguished below.

In Cases A and B, $R$ may fall below 1, indicating the quantum strategy should be chosen depending on the $N$ available. The region at very low $N$ where simultaneous estimation is superior may be dismissed as this is a very short range of $N$ where the error is high. However, there are then significant regions where the individual, and then the simultaneous, strategies are preferred, before the classical strategy takes over at high $N$. In Case C, the first minimum of $R$ is above 1, so the simultaneous estimation scheme always outperforms the individual, until it is overtaken by the classical scheme.

We analyze the optimal attainable error $\Delta_{\min}$ and the optimal size $N_{\mathrm{opt}}$ of the initial GHZ state that produces this error. Of course with high $N$ one would always use the classical strategy with as many probe qubits as possible. We therefore look at the more interesting regime where $N$ is limited and the quantum strategies display an advantage. Incidentally, this regime is also useful at large $N$ as a quantum enhanced technique may be used at low $N$ and then repeated `classically' to improve the precision. Hence the initial minima of the errors of the two quantum regimes are compared. In Case A, the minimum is provided by the individual strategy (i.e., $\Delta_{\min}=\Delta_{\mathrm{ind}}(2N_{\mathrm{opt}})$) and in Cases B and C it is provided by the simultaneous strategy (i.e., $\Delta_{\min}=\Delta_{\mathrm{sim}}(2N_{\mathrm{opt}})$). Figure~\ref{fig:FullContourPlot} (Left) shows the optimal value $N_{\mathrm{opt}}$ of $N$ that gives the variation of this minimum error with $\varphi$ and $\kappa$ as well as the boundaries that separate the behavioral regimes, whilst Figure~\ref{fig:FullContourPlot} (Right) depicts the variation of the minimum error itself $\Delta_{\min}$ with the parameters. One can see that Case A manifests at low $\varphi$ and high $\kappa$ (i.e.~low noise); this is followed by Case B, and  then C as the noise increases with decreasing $\kappa$ and increasing $\varphi$.

\section{Trade-off between sequential and parallel entangled schemes: The role of $M$}\label{sec:roleofM}

\begin{figure}
    \centering
    {
        \includegraphics[width=7cm]{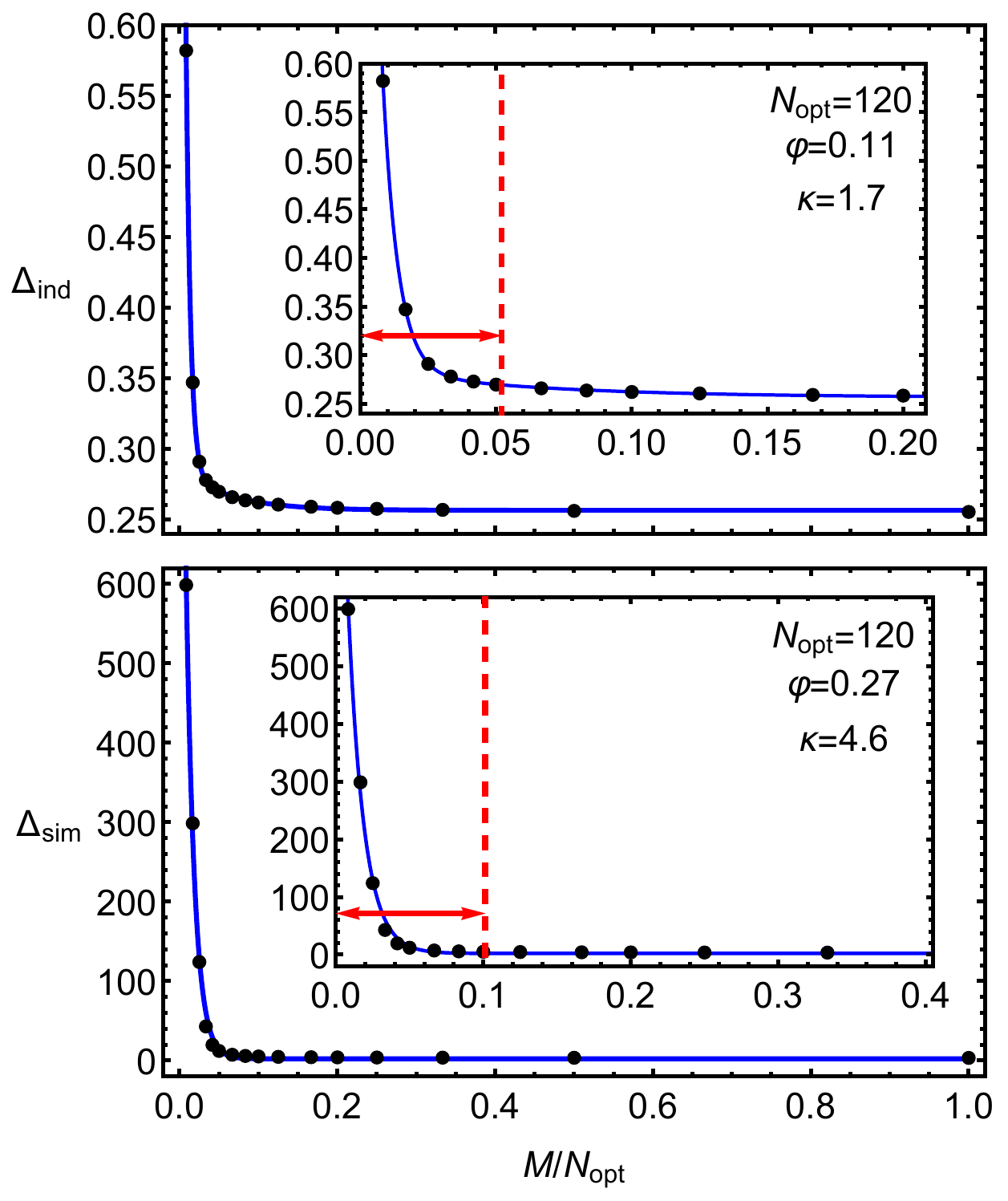}
    }
    \caption{(color online) The estimation error against $M/N_{\mathrm{opt}}$ for the metrological scheme of Figure~\ref{interpolate}.
    	Top: An example of the behavior of the precision in the case the individual estimation scheme is optimal. Here, $\varphi = 0.11$ and $\kappa = 1.7$, which gives $N_\mathrm{opt} = 120$. The precision falls within 5\% of that of a GHZ state of $N_\mathrm{opt}$ qubits after using a GHZ state of only 6 qubits (5\% of $N_\mathrm{opt}$).
    	Bottom: Here, the simultaneous scheme is optimal and $\varphi = 0.27$ and $\kappa = 4.6$ are used, again giving $N_\mathrm{opt} = 120$. The precision reaches 5\% of that of the $N_\mathrm{opt}$-qubit GHZ state after a GHZ state of only 12 qubits (10\% of $N_\mathrm{opt}$) is used. All the plotted quantities are dimensionless.}
    \label{RoleofM}
\end{figure}

In the previous section, we have shown that the optimal size of an entangled probe state, $M=N_{\mathrm{opt}}$, varies with the values of the parameters and may be in principle very large.
Experimentally, creating entangled states with a large number of particles is a costly process, limited by technological constraints.
Here, we demonstrate that the limit of using large entangled states may be overcome in principle by utilizing a framework which is a hybrid of the sequential and parallel schemes traditionally studied \cite{Giovannetti2006}, as illustrated in Figure~\ref{interpolate}, and choosing a suitable $M \ll  N_{\mathrm{opt}}$.
In this case, the noisy phase imprinting operation alters the probe state as
\begin{equation}
\rho=\left(\Lambda^{N_{\mathrm{opt}}/M}\right)^{\otimes M}[\rho_{0}],
\end{equation}
where $\rho_{0}$ is now an $M$-qubit GHZ state and $\Lambda^{N_{\mathrm{opt}}/M}$ can be obtained by replacing $b$ and $c$ by $B=\frac{1}{2}\left(1-(1-2b)^{N_{\mathrm{opt}}/M}\right)$ and $c^{N_{\mathrm{opt}}/M}$ in Eq.~(\ref{Map's matrix}), respectively.
As in Eq.~(\ref{twoparts}) before, the output state can be divided into diagonal and non-diagonal parts as
\begin{eqnarray}
\nonumber  \varrho_{1} &=& \frac{1}{2}\left( \left(B^{M}+(1-B)^{M}\right)\left(|0\rangle\langle0|^{\otimes M}+|1\rangle\langle1|^{\otimes M}\right)\right.\\
\nonumber &+ &\left. c^{N_{\mathrm{opt}}}|0\rangle\langle1|^{\otimes M}+c^{\ast N_{\mathrm{opt}}}|1\rangle\langle0|^{\otimes M} \right),\\
\nonumber \varrho_{2}&=&\frac{1}{2}\sum_{m=1}^{M-1}\left(B^{m}(1-B)^{M-m}+B^{M-m}(1-B)^{m}\right)\\
& \times& |0\rangle\langle0|^{\otimes M-m} \, |1\rangle\langle1|^{\otimes m}.
\end{eqnarray}
By inspecting the estimation precision over a range of $M$, we can see that $N_{\mathrm{opt}}/M$ sequential channel applications on each qubit of a GHZ state of only a few qubits may achieve metrological performances very close to that of a full GHZ state of $N_{\mathrm{opt}}$ qubits.
By scanning the parameter range, we can verify that this behavior is independent of the parameter values and the strategy implemented, so that this result is generally valid across our investigation.
This is demonstrated in Figure~\ref{RoleofM} where the estimation precision versus $M/N_{\mathrm{opt}}$ is plotted. We find that the error shrinks exponentially with increasing $M/N_{\mathrm{opt}}$ and only a very small percentage of $N_{\mathrm{opt}}$ is required to saturate the precision.
As quantum states with an increasing number of qubits are routinely engineered \cite{ghzexp}, the experimental realization of our multiparameter sensing may become feasible in the near future.

\section{Conclusions}\label{sec:Conclusions}

We investigated resources and strategies in multiparameter quantum metrology, focusing on a physical model of phase estimation where the generator of the phase shift is randomly sampled according to a distribution with concentration parameter $\kappa$ \cite{Rosanna}. We  explored how to estimate precisely and efficiently both $\kappa$ and the phase $\varphi$, having the availability of a finite number $N$ of interactions with the phase imprinting channel. We adopted a performance ratio $R$ as figure of merit to account for the advantage, in terms of reduction of resources up to a factor two, of estimating the parameters simultaneously as opposed to individually, and examined  compatibility of such multiparameter estimation depending on the chosen probe states following the analysis in \cite{Ragy}. While for single qubit and two qubit probes compatibility is not always fully met, we find that simultaneous estimation is always advantageous. We assessed the role of multipartite entanglement by developing strategies for the considered problem using GHZ probe states of up to $N$ qubits, and providing concrete recipes to achieve quantum enhanced estimation at finite $N$ using either individual or simultaneous strategies, depending on the parameter range. Crucially, we showed that large $N$-partite entangled resources are not needed for this enhancement, as a very small portion $M \ll N$ of entangled qubits suffices to match such performance by suitably distributing the $N$ channel applications in a probing geometry which interpolates between the conventional sequential and parallel metrological schemes (see Figure~\ref{interpolate}). Finally, at large $N$, no quantum enhancement survives and a classical strategy based on $N$ independent repetitions of single qubit estimation always attains optimality.

Our analysis highlights several interesting features with potential relevance for practical applications to sensing technologies. Schemes such as the hybrid one in Figure~\ref{interpolate} can be seen as a {\em compression} of the fully parallel strategy which drastically reduces the resources without any noticeable degradation in precision. Such a compression could be applied to many existing implementations of sensing and metrology \cite{Giovannetti} using e.g.~cold atoms, photonic qubits, or nitrogen vacancy centres in diamond. It would be worthwhile to develop general methods to optimize the circuit architecture given any specific noisy quantum metrology setting arising in applications, possibly exploiting semidefinite programming techniques as in \cite{chiribella}.

Our model realizes an instance of a unital phase-covariant channel, as illustrated in Figure~\ref{fig1}. In the future, it would be interesting to extend our analysis to general non-unital phase-covariant channels, and investigate joint estimation of a phase $\varphi$ and of all the parameters specifying the noise (the deformations $\lambda_\parallel$ and $\lambda_\perp$ as well as a Bloch displacement vector).

Finally, while we considered more general states for single and two qubit settings, we have specified GHZ states as initial probe states to analyze strategies relying on $N$-partite entanglement, partly due to their structural simplicity and the fact that they satisfy the compatibility condition formalized in Eq.~(\ref{compa2}). However, these states may not be optimal in general instances of noisy quantum metrology  \cite{noisy freqency estimation,Ragy}. A notable extension could be to develop efficient algorithms to identify optimal probe states in multiparameter quantum estimation, possibly extending the methods of \cite{paolo}.

\begin{acknowledgments}
We acknowledge discussions with T.~R.~Bromley, L.~A.~Correa, A.~Gorecka, P.~A.~Knott, K.~Macieszczak, K.~Modi, A.~Sorouri.
RN and GA acknowledge funding from the European Research Council (ERC) Starting Grant GQCOP (Grant No.~637352),  the Foundational Questions Institute (fqxi.org) Physics of the Observer Programme (Grant No.~FQXi-RFP-1601), and the Royal Society International Exchanges (Grant No.~IE150570).
\end{acknowledgments}


\begin{thebibliography}{99}
\bibitem{Giovannetti2006} V. Giovannetti, S. Lloyd, and L. Maccone, Phys. Rev. Lett. \textbf{96}, 010401 (2006);


\bibitem{Braun}D. Braun, et al., arXiv:1701.05152 (2017).
\bibitem{Pezze}  L. Pezz\`e, et al., arXiv:1609.01609 (2016).

\bibitem{Giovannetti} V. Giovannetti, S. Lloyd, and L. Maccone, Science \textbf{306}, 1330 (2004); V. Giovannetti, S. Lloyd, and L. Maccone, Nat. Photon. \textbf{5}, 222 (2011).

\bibitem{navigation} K. Bongs, R. Launay, and M. A. Kasevich, Appl. Phys. B \textbf{84}, 599 (2006).

\bibitem{biology} P. M. Carlton, J. Boulanger, C. Kervrann, J.-B. Sibarita, J. Salamero, S. Gordon-Messer, D. Bressan, J. E. Haber, S. Haase, L. Shao, et al., Proc. Natl. Acad. Sci. \textbf{107}, 16016 (2010); M. A. Taylor, J. Janousek, V. Daria, J. Knittel, B. Hage, H.-A. Bachor, and W. P. Bowen, Nature Photon. 7, \textbf{229} (2013).

\bibitem{gravwaves} The LIGO Scientific Collaboration, Nat. Phys. \textbf{7} 962 (2011); J. Aasi et al. Nat. Photon. \textbf{7} 613 (2013)


\bibitem{like phase estimation} S. Simmons, et al. Phys. Rev. A \textbf{82}, 022330 (2010); A. Crespi, et al., Appl. Phys. Lett. \textbf{100}, 233704 (2012); M. Tsang, New J. Phys. {\bf 15}, 073005 (2013).

\bibitem{coherence} A. Streltsov, G. Adesso, and M. B. Plenio, arXiv:1609.02439 (2016).

\bibitem{noisy freqency estimation} R. Demkowicz-Dobrza\'{n}ski, et al., Nat. Commun. \textbf{3}, 1063 (2012); J. Ko{\l}ody\'{n}ski, and R. Demkowicz-Dobrza\'{n}ski, New J. Phys. \textbf{15}, 073043 (2013); B. M. Escher, et al., Nat. Phys. \textbf{7}, 406 (2011); A. Smirne, et al., Phys. Rev. Lett. \textbf{116}, 120801 (2016); S. F. Huelga, et al., Phys. Rev. Lett. \textbf{79}, 3865 (1997); Y. Matsuzaki, et al., Phys. Rev. A \textbf{84}, 012103 (2011); A. W. Chin, et al., Phys. Rev. Lett. \textbf{109}, 233601 (2012).

\bibitem{Yousefjani} R. Yousefjani, S. Salimi, A. S. Khorashad, arXiv:1508.01990 and arXiv:1602.01691.

\bibitem{lossy interferommetry} U. Dorner, et al., Phys. Rev. Lett. \textbf{102}, 040403 (2009); R. Demkowicz-Dobrza\'{n}ski, et al., Phys. Rev. A \textbf{80}, 013825 (2009); M. Kacprowicz, et al., Nat. Photon. \textbf{4}, 357–360 (2009).

\bibitem{macconehierarchy} R. Demkowicz-Dobrza\'{n}ski and L. Maccone, Phys. Rev. Lett. ´
{\bf 113}, 250801 (2014); P. Sekatski, et al., arXiv:1603.08944 (2016).

\bibitem{Rosanna} R. Nichols, T. R. Bromley, L. A. Correa, and G. Adesso, Phys. Rev. A \textbf{94}, 042101 (2016).

\bibitem{Fisher inf. Matrix} C. W. Helstrom, \textit{Quantum Detection and Estimation Theory} (Academic, New York, 1976);
S. L. Braunstein and C. M. Caves,
Phys. Rev. Lett. {\bf 72}, 3439 (1994); M. G. A. Paris, Int. J. Quantum Inform. \textbf{7}, 125 (2009).


\bibitem{Review} M. Szczykulska, T. Baumgratz, and A. Datta, Adv. Phys.: X \textbf{1}, 621 (2016).

\bibitem{Ragy} S. Ragy, M. Jarzyna, and R. Demkowicz-Dobrza\'{n}ski, Phys. Rev. A \textbf{94}, 052108 (2016).

\bibitem{magfields} L. M. Pham, D. Le Sage, P. L. Stanwix, T. K. Yeung, D. Glenn, A. Trifonov, P. Cappellaro, P. Hemmer, M. D. Lukin, H. Park, et al., New Journal of Physics \textbf{13}, 045021 (2011);  T. Baumgratz and A. Datta, Phys. Rev. Lett. \textbf{116}, 030801 (2016).

\bibitem{Liu} J. Liu, X.-X. Jing, and X. Wang, Sci. Rep. \textbf{5}, 8565 (2015).

\bibitem{TsangX}
D. W. Berry, M. Tsang, M. J. W. Hall, and H. M. Wiseman, Phys. Rev. X {\bf 5}, 031018 (2015).


\bibitem{two noise}
A. Fujiwara and H. Imai, J. Phys. A: Math. Gen. {\bf 36}, 8093 (2003);
A. Monras and F. Illuminati, Phys. Rev. A \textbf{83}, 012315 (2011).

\bibitem{unitary parameters}
M. A. Ballester, Phys. Rev. A {\bf 69}, 022303 (2004);
M. A. Ballester, Phys. Rev. A {\bf 70}, 032310 (2004);
M. G. Genoni, et al., Phys. Rev. A {\bf 87}, 012107 (2013);
C. Vaneph, T. Tufarelli, and M. G. Genoni, Quant. Meas. Quant. Metrol. {\bf 1}, 12 (2013);
P. C. Humphreys, M. Barbieri, A. Datta, and I. A. Walmsley, Phys. Rev. Lett. \textbf{111}, 070403 (2013); J.-D. Yue, Y.-R. Zhang, and H. Fan, Sci. Rep. \textbf{4}, 5933 (2014); J. Liu, X.-M. Lu, Z. Sun, and X. Wang, J. Phys. A: Math. Theor. {\bf 49}, 115302 (2016); T. Baumgratz and A. Datta, Phys. Rev. Lett. \textbf{116}, 030801 (2016); C. N. Gagatsos, D. Branford, and A. Datta,
Phys. Rev. A {\bf 94}, 042342 (2016); M. A. Ciampini, et al., Sci. Rep. {\bf 6}, 28881 (2016);
P. A. Knott, et al., Phys. Rev. A {\bf 94}, 062312 (2016);
P. Kok, J. Dunningham, and J. F. Ralph,
Phys. Rev. A {\bf 95}, 012326 (2017); K. Duivenvoorden, B. M. Terhal, and D. Weigand,
Phys. Rev. A {\bf 95}, 012305 (2017);
T. J. Proctor, P. A. Knott, and J. A. Dunningham, arXiv:1702.04271 (2017).


\bibitem{unitary and noise} S. I. Knysh, and G. A. Durkin, arXiv:1307.0470; M. D. Vidrighin, G. Donati, M. G. Genoni, X.-M. Jin, W. S. Kolthammer, M. S. Kim, A. Datta, M. Barbieri, and I. A. Walmsley, Nat. Commun. \textbf{5}, 3532 (2014); P. J. D. Crowley, A. Datta, M. Barbieri, and I. A. Walmsley, Phys. Rev. A \textbf{89}, 023845 (2014).



\bibitem{interpower}
D. Girolami, et al., Phys. Rev. Lett. {\bf 112}, 210401 (2014); A. Farace, et al., New J. Phys. {\bf 18}, 013049 (2016).



\bibitem{covariant channel} A. S. Holevo, Rep. Math. Phys \textbf{32}, 211 (1993); J. Math. Phys. \textbf{37}, 1812 (1996). I. Bengtsson and K. Zyczkowski, \textit{Geometry of Quantum States: An Introduction to Quantum Entanglement} (Cambridge University Press, Cambridge, 2006).

\bibitem{von Mises-Fisher} R. Fisher, Proc. R. Soc. Lond. A \textbf{217}, 295 (1953).

\bibitem{Convexity} A. Fujiwara, Phys. Rev. A \textbf{63}, 042304 (2001).

\bibitem{QFI for states in exponential form} Z. Jiang, Phys. Rev. A \textbf{89}, 032128 (2014).

\bibitem{GHZ}
D. M. Greenberger, M. A. Horne, A. Shimony, and A. Zeilinger, Am. J. Phys.
{\bf 58}, 1131 (1990).

\bibitem{ghzexp}
T. Monz, et al., Phys. Rev. Lett. {\bf 106}, 130506 (2011).

\bibitem{chiribella} G. Chiribella, New J. Phys. {\bf 14} 125008 (2012).

\bibitem{paolo}
P. A. Knott, New J. Phys. {\bf 18}, 073033  (2016)

\end{thebibliography}
\end{document}